\documentclass{cimento}
\usepackage{graphicx}  
\usepackage{enumitem}
\usepackage{graphbox}

\title{SiPMs for cryogenic temperature}
\author{I.~Kochanek for DarkSide Collaboration}
{\instlist{\inst Laboratori Nazionali del Gran Sasso}}

\begin{document}

\maketitle

\begin{abstract}
The DarkSide-20k collaboration is preparing to equip $20~m^2$ of SiPMs working in liquid argon at 86~K for the direct search of WIMPs. 
The collaboration had to solve many technological aspects, such as the development of SiPM optimized for operation in liquid argon, 
the readout of large SiPM-based detectors, the reliable packaging of more than 200000 SiPMs using radiopure materials. 
The packaging solutions available for cryogenic applications and the performances of the newest cryogenic extended gain SiPMs from FBK will be discussed.
\end{abstract}

\section{DarkSide-20k}
DarkSide is a stepped program for the realization of increasing size dark matter experiments toward the direct detection of WIMP interactions on a liquid argon target.
DarkSide-50 is based on a two-phase Time Projection Chamber (LAr TPC) with a fiducial mass of 47~kg of Underground Liquid Argon (UAr).
The underground origin of argon, extracted from a CO$_2$ well in Colorado, shows a depletion of $\,^{39}$Ar, which represent the highest background for argon-based detectors, in excess of 1400. 
DarkSide-50 is inserted in an active Liquid Scintillator Veto (LSV) inside a cylindrical Water Cherenkov Veto (WCV) to further reduce the background. 
The experiment is installed at Laboratori Nazionali del Gran Sasso, under 1400~m of overburden rock providing shield against cosmic rays. 
DarkSide-50 demonstrated exceptional results in the WIMP search, thanks to the extremely low background achieved \cite{ref:low}, \cite{ref:532-days}.

Particle interactions in the active volume produce a scintillation light (S1) and ionization electrons. 
Electrons are drifted by means of an electric field to the liquid-gas interface, extracted in the gas phase.
There the accelerated electrons produce a second light pulse (electroluminescence), called S2.
Photons from S1 and S2 are detected by 38 3" photo-multipliers (PMTs) installed on top and bottom of the TPC.
The presence of the S1 and S2 allows to reconstruct the position of the original interaction with centimeter resolution.
The discrimination of nuclear recoils from electromagnetic interaction is performed by a Pulse Shape Discrimination (PSD) of the S1 scintillation with a rejection better than $10^8$ \cite{ref:532-days}.

DarkSide-20 is the next installation and will be based on a scaled design with an active (fiducial) mass of 23 t (20 t) \cite{ref:ds20k}. 

\section{Light detection in DarkSide-20k}
Accurate light detection is of primary importance for the DarkSide program.
The experience from DarkSide-50, where the collaboration was forced to run the PMTs at very low bias and to adopt cryogenic low-noise preamplifier, pushed to start an intense R\&D program for the development of innovative silicon-photomultipliers (SiPM) based photodetectors.

\begin{figure}[b]
  \begin{center}
    \includegraphics[trim={100 150 20 5cm}, clip, height=0.2\textheight]{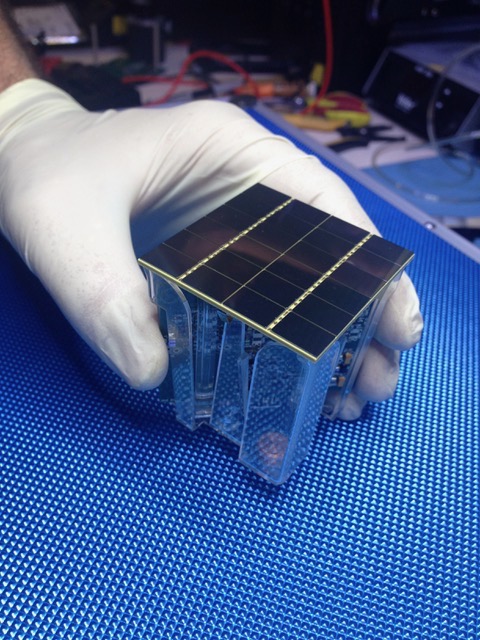}     
    \hskip 2cm
    \includegraphics[trim={200 10 200 22cm}, clip, height=0.2\textheight]{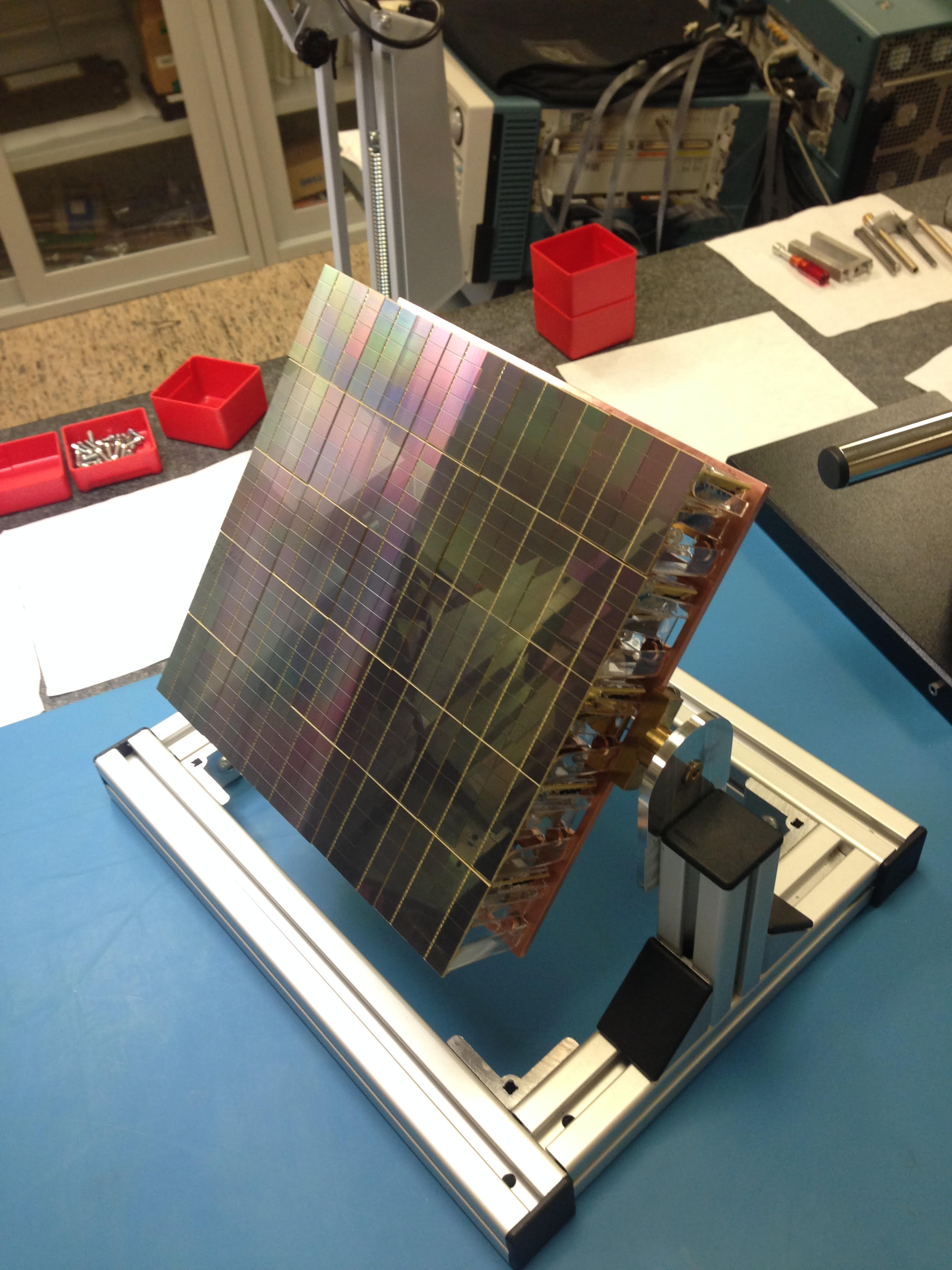}     
    \caption{A Photo Detector Module, which includes a tile with 24 mounted SiPMs and a front-end board in an acrylic cage (left).
    A motherboard with mounted 25 PDMs (right).}
    \label{fig:pdm}
  \end{center}
\end{figure}

A SiPM is constituted by a large collection of parallel SPADs (Single Photon Avalanche Diode) of typical size $20-50~\mathrm{\mu m}$.
Each SPAD operates independently in Geiger mode.
A signal is generated when few SPADs are triggered by incoming photons or by dark-rate. 
SiPMs typically exhibit an extraordinary charge resolution O(1\%), at least an order of magnitude better than PMTs. 
Furthermore, the photon detection efficiency, which includes both the quantum efficiency and the SPAD fill factor, can read 60\% on the full detector surface.
On the other hand, the dark-rate from SiPMs lays around $10^4 - 10^5~\textrm{cps/mm}^2$ at 300~K, with the addition of correlated noises of about 15-20\%. 

SiPMs are small size devices (tens to hundreds of mm$^2$), while for DarkSide-20k a 3" PMT replacement is required.
In DarkSide this is obtained by grouping SiPMs in arrays of 25~cm$^2$, called \textit{tiles}, with a cryogenic front-end board that pre-amplifies and aggregates all the signal to a single analog output. 
The challenge is to reliably assemble a Photo Detector Module (PDM), see Figure~\ref{fig:pdm} left, capable of operating at 87~K with a signal to noise ratio (defined as gain divided the baseline noise) larger than 8 and a dark rate smaller than $80~\textrm{mcps/mm}^2$.
Up to now, SiPMs are produced by FBK (Fondazione Bruno Kessler) developed in collaboration with DarkSide for low after pulsing and low dark rate operations in liquid argon.

For low background detectors, a high radio-purity of all the components building the detector is required. 
SiPMs are made of high purity silicon and are intrinsically radio-pure, however a PDM includes electronic components and connectors. 
A second challenge for the collaboration is to identify components and procedures that will maintain the contaminations of U/Th at the level of mBq/PDM.

Finally in Darkside-20K, about 10000 PDMs are foreseen.
This requires the use of highly scalable packaging techniques and the deployment of a industrial-grade packaging facility capable of supporting our requirements in term of reliability and radio-cleanliness.

The PDMs are installed in a 25 units structure called {\it motherboard} (MB) providing interface to the TPC. 
The final goal for radio-purity is to reach around 10~mBq/MB.

In October 2018 we produced the first full motherboard. 
The next step is to move the production of SiPM to a commercial foundry that can provide higher production throughput and better repeatability.
INFN selected LFoundry. 
However, before starting the large scale production we need to finalize the design parameters of the SiPMs.
This is performed in a dedicated setup at LNGS capable of characterizing the dark-rate and the correlated noises as function of temperature in the range 40 to 300~K \cite{ref:cryo}. 
In the same setup it is possible to further optimize the design of the front-end electronics to match at best the parameters of the light detectors.

\section{Packaging solutions}
\begin{table}[t]
  \caption{CTE values and radio-cleaness in U/Th/K chains for common substrate materials.}
  \footnotesize
  \label{tab:cte}
  \begin{tabular}{llll}
    \hline
    Material & CTE $[$ppm/K$]$ & U/Th $[$mBq/kg$]$ & $^{40}$K $[$mBq/kg$]$ \\
    \hline
    Silicon & 3-5 & very low & very low \\
    FR4 & 12-14 & very high & very high \\
    Polyimide/Kapton & 20 & 1-10 & $>$10 \\
    Arlon NT & 5-7 & 100 & 1000 \\
    Selected Fused Silica & 0.5 & 0.05 & - \\
    \hline
  \end{tabular}
\end{table}

For packaging the tiles of DarkSide-20k several aspects have to be taken under considerations:
\begin{itemize}[nosep]

\item The Thermal Expansion Coefficient (CTE) of the silicon is quite low, a compatible substrate is required to avoid mechanical stress on the silicon die.
From Table~\ref{tab:cte}, it is clear that the best solution for the substrate would be silicon itself.
Silicon PCBs are not trivial to make, considering the requirement of insulation, stray capacitance (between traces and substrate) and the requested low impedance ground plane 
(to limit the noise pickup).
Arlon 55NT is the baseline solution.
An R\&D is ongoing on multilayer hybrid fused silica-polimide PCB. 
This alternative would significantly reduce the neutron background from U/Th $\alpha$-n production.
The second contribution in background is from connectors, where glass-fiber reinforced nylon exhibits significant contaminations. 
Electronic components are at the level of 60-100~$\mu$Bq/kg.

\item The current bonding procedure uses conductive cryogenic epoxy EPO-TEK EJ2189. 
The epoxy is applied by a robot dispenser, Nordson EFD, with nine dots pattern per SiPM. 
Then the SiPMs are positioned with a manual die bonder, Westbond 7200CR. 
Finally the tiles are cured for one hour at 100$^{\circ}$C.
The conductive epoxy is required because the backside contact of the SiPM is TTiN based and no direct solder is possible on this finish.
For the first motherboard we bonded more than 600 dies and none showed any problem in liquid nitrogen, leading to a lower limit on the packaging reliability of $>$99.8\%.
However, the silver loaded epoxy is not radio-clean and we have no data on long term stability.
A new procedure that we are working on, involves indium/solder (eutectic) bump bonding or gold studs ball bonding.
Both solutions were tested and provided very strong bonds. 
LFoundry is currently producing dies with a copper-based backside contact pads that will be used to fine tune the bonding procedure.

\item The anode contact is on the top side of the SiPM and, up to now, it is connected via a wire bonding. 
We use 25~$\mu$m aluminium wire, 24 bonds per PDM.
This makes the manipulation of the tiles very complicated.
Since we want to optimise the detector fill factor there is no dead space around the tile and the silicon surface can not be touched to avoid damaging the wire bondings. 
At the same time identifying broken wires requires a characterization of the tile on a test bench.
This results in a very complicated detector installation.
A better solution would be provided by the use of Through Silicon Vias (TSV). 
We are in contact with LFoundry that is developing TSVs for SiPMs according to our specifications.
The bump bonding technique that we plan to use is already compatible with TSVs.
\end{itemize}

\section{Tiles characterization}
\begin{figure}[t]
  \begin{center}
    \includegraphics[width=0.45\textwidth]{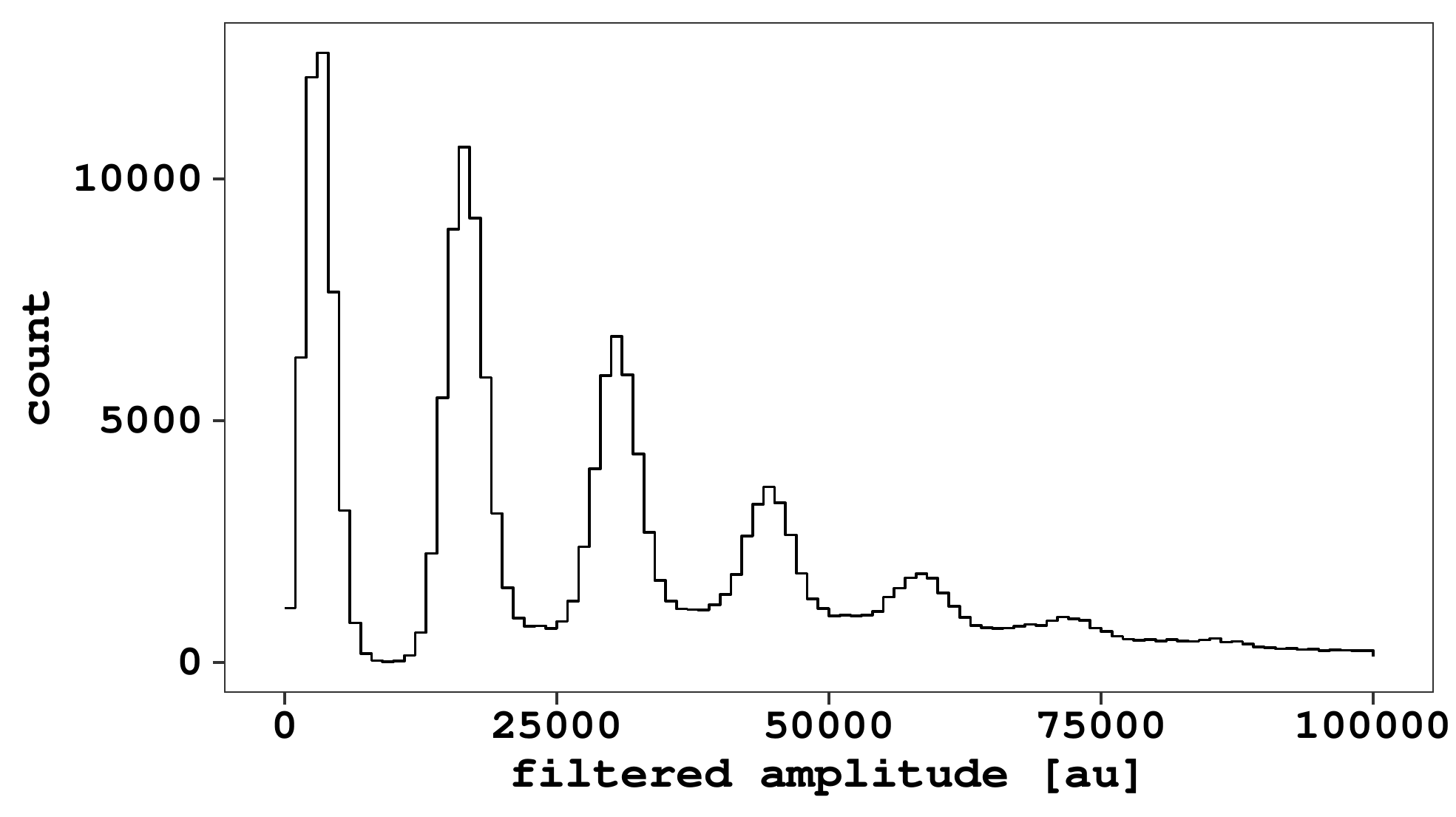}     
    \includegraphics[width=0.45\textwidth]{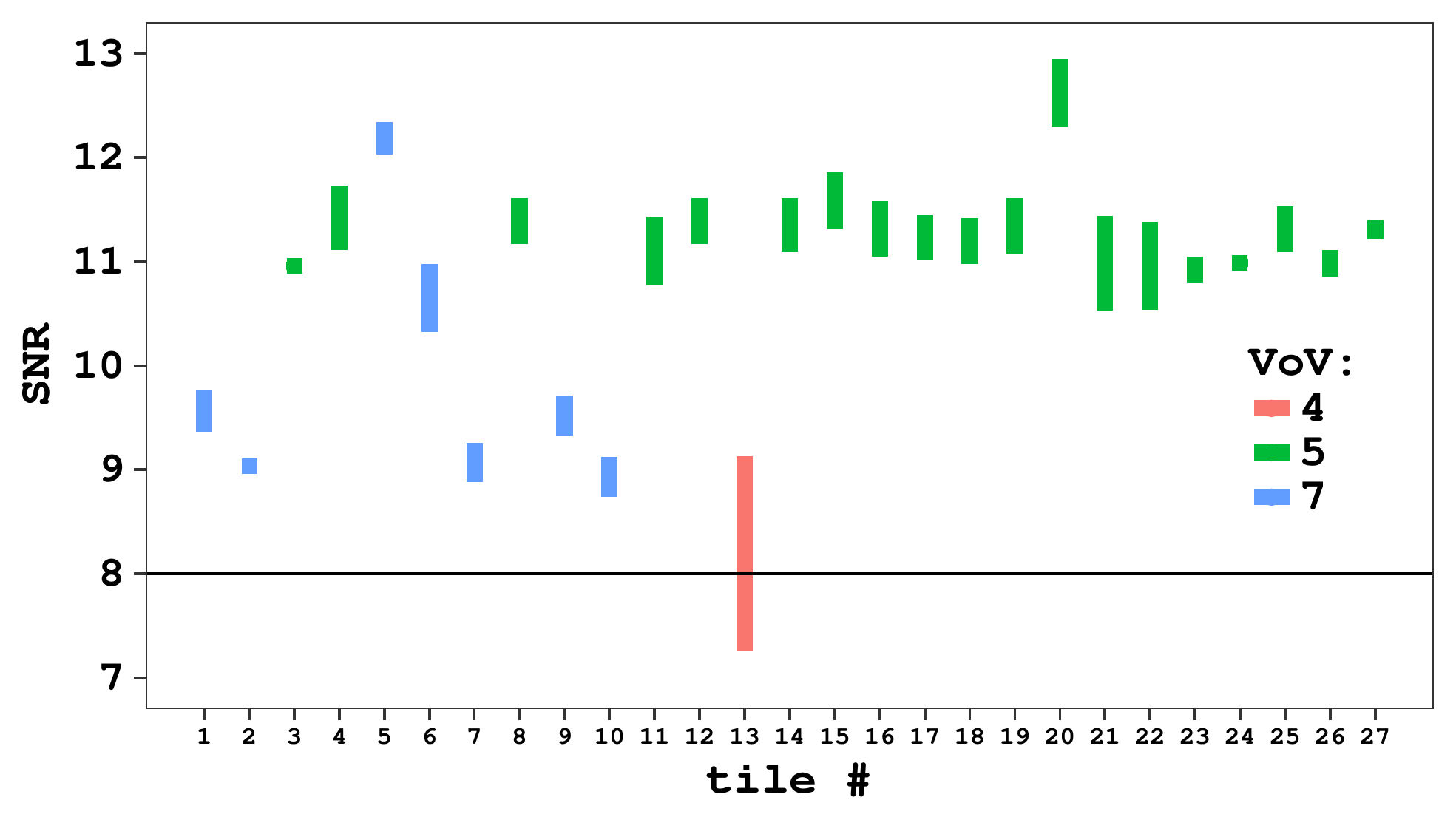}     
    \caption{Single photoelectron peaks (left) and signal to noise distribution for tested tiles (right).}
    \label{fig:pe}
  \end{center}
\end{figure}

For the production of the first motherboard we produced more than 30 PDMs. 
We established a rigorous test procedure to qualify them.
The procedure includes tests at room temperature and in liquid nitrogen. 
The tests are targeted to verify that all 24 SiPMs of each tile are working according to our specifications and that the front-end board is operating as expected. 
Finally we acquired the waveform of the photoelectrons produced by a laser.
The analysis (based on advanced digital signal processing algorithms) showed that for all the PDMs the photoelectron peaks are well distinguishable with a signal to noise ratio above 8, Figure~\ref{fig:pe}.

\section{Conclusions}
DarkSide-20k succeeded in an ambitious program to deploy radio-pure cryo-graded, SiPM-based, large photodetectors.
The development included R\&D in SiPM at FBK, in extremely low noise preamplifier board at LNGS, in the most avanced silicon packaging techniques at the Princeton University.
The next step is the realization of a cutting edge silicon packaging facility at LNGS, called Nuova Officina Assergi (NOA) where DarkSide and other experiments can build
their future photo-detectors.


\begin{thebibliography}{0}
  \bibitem{ref:low} \BY{DarkSide Collaboration}
    \IN{Phys. Rev. Lett.}{121}{2018}{081307};
  \bibitem{ref:532-days} \BY{DarkSide Collaboration}
    \IN{Phys. Rev. D}{98}{2018}{102006};
  \bibitem{ref:ds20k} \BY{DarkSide Collaboration}
    \IN{The European Physical Journal Plus}{133}{2018}{131};
  \bibitem{ref:cryo} \BY{F. Acerbi, et. al}
    \IN{IEEE Transactions on Electron Devices}{64, 2}{2017}{16618176};

\end{thebibliography}
\end{document}